\documentclass[aip,
apl
 amsmath,amssymb,
 reprint,%
]{revtex4-2}

\usepackage{graphicx}
\usepackage{dcolumn}
\usepackage{bm}
\usepackage{xcolor}

\usepackage{multirow}

\usepackage{booktabs}
\usepackage{enumitem}
\usepackage{textcomp}

\usepackage[utf8]{inputenc}
\usepackage[T1]{fontenc}
\usepackage{mathptmx}
\usepackage{etoolbox}

\newcommand\red[1]{\textcolor{red}{#1}}
\newcommand\blue[1]{\textcolor{blue}{#1}}

\usepackage[normalem]{ulem}
\usepackage{ifthen}
\usepackage{xspace}
\newcounter{showSGedits}
\newcommand\FD[1]{\ifthenelse{\value{showSGedits}=0}{#1}{\textcolor{teal}{#1}}}
\newcommand\rsout[1]{\ifthenelse{\value{showSGedits}=0}{}{\red{\sout{#1}}}}
\newcommand\brep[1]{\ifthenelse{\value{showSGedits}=0}{#1}{\blue{#1}}}
\newcommand{\mnote}[1]{\ifthenelse{\value{showSGedits}=0}{}{\marginpar{\scriptsize \red{#1}}}}
\newcommand{\mnoteFD}[1]{\ifthenelse{\value{showSGedits}=0}{}{\marginpar{\scriptsize \FD{#1}}}}

\makeatletter
\def\@email#1#2{%
 \endgroup
 \patchcmd{\titleblock@produce}
  {\frontmatter@RRAPformat}
  {\frontmatter@RRAPformat{\produce@RRAP{*#1\href{mailto:#2}{#2}}}\frontmatter@RRAPformat}
  {}{}
}%
\makeatother
\begin{document}

\preprint{AIP/123-QED}

\title{Low two-level-system noise in hydrogenated amorphous silicon}

\author{Fabien Defrance}
\email[]{fabien.m.defrance@jpl.nasa.gov}
\affiliation{Jet Propulsion Laboratory, California Institute of Technology, Pasadena, CA 91109, USA}

\author{Andrew D. Beyer}
\affiliation{Jet Propulsion Laboratory, California Institute of Technology, Pasadena, CA 91109, USA}

\author{Jordan Wheeler}
\affiliation{National Institute of Standards and Technology, Boulder, CO 80305, USA}

\author{Jack Sayers}
\affiliation{California Institute of Technology, Pasadena, CA 91125, USA}

\author{Sunil R. Golwala}
\affiliation{California Institute of Technology, Pasadena, CA 91125, USA}

\date{\today}

\begin{abstract}
At sub-\rsout{k}\brep{K}elvin temperatures, \rsout{an excess frequency noise is generated by}two-level\rsout{-}\brep{ }systems (TLS\rsout{s}) 
\rsout{defects existing inside}\brep{present in} amorphous dielectrics \brep{source a permittivity noise}, degrading the performance of a wide range of devices using superconductive resonators such as qubits or kinetic inductance detectors.  
We report here on measurements of TLS noise in hydrogenated amorphous silicon (a-Si:H) films deposited by plasma-enhanced chemical
vapor deposition (PECVD) in \FD{superconductive} lumped-element resonators using parallel-plate capacitors \brep{(PPCs)}.  
The TLS noise results presented in this article for two recipes of a-Si:H improve on the best achieved in the literature by a factor >5 for a-Si:H and other amorphous dielectrics and are comparable to \rsout{that}\brep{those} observed 
\rsout{with crystalline dielectrics}\brep{for resonators deposited on crystalline dielectrics}.

\vspace{6mm}

\copyright  2024. All rights reserved.
\end{abstract}

\maketitle



\FD{Superconductive}
\rsout{detectors}\brep{devices}
like kinetic inductance detectors (KIDs) and superconductive qubits exhibit an excess frequency noise and loss at low temperatures (below a few Kelvins) that was shown~\cite{Gao:2007, Gao:2008c, Gao:2008b} to originate from 
\rsout{two-level-system (TLS) defects}\brep{two-level systems (TLS)}~\cite{Phillips:1987, Esquinazi:1998}.  
The standard tunneling model (STM) describes 
TLS\rsout{s} as
\rsout{atoms or groups of atoms that can oscillate between two different lattice sites}\brep{defects that can switch between two different configurations}\cite{Muller:2019, Phillips:1987}\rsout{,}\brep{.} 
\rsout{resulting in fluctuations of TLS electric dipole moments and thus the material's permittivity and the resonator's resonant frequency}\brep{Such TLS can also change state by phonon emission, introducing a loss mechanism that, via the fluctuation-dissipation theorem\cite{Callen:1951}, results in noise in dielectric permittivity and thus, in resonators, resonant frequency}~\cite{Gao:2007, Gao:2008c, Gao:2008b, Noroozian:2009, Zmuidzinas:2012}. 
\rsout{Because crystalline dielectrics lack dangling bonds and exhibit reduced atomic-scale disorder, assumed to cause TLSs in amorphous dielectrics,}\brep{Because the atomic scale disorder giving rise to TLS in amorphous dielectrics is vastly reduced in crystalline dielectrics,}
they are preferentially selected as substrates for 
\rsout{low-noise and low-loss superconducting resonators.}\brep{superconductive devices where low loss and low noise are critical, though it should be noted that unavoidable surface oxides still host TLS\cite{Gao:2008b}.}
However, fabricating multi-layer structures with crystalline dielectrics remains extremely challenging~\cite{Denis:2009, Beyer:2017}, 
\rsout{which currently constrains}\brep{constraining}
such devices to single-layer architectures \brep{on crystalline substrates}.
\rsout{This limitation can be very constraining in components like KIDs, where single layer interdigitated capacitors (IDCs) occupy a larger footprint on a telescope focal plane (where real estate is often very limited and expensive) than 2-layer parallel plate capacitors (PPCs), and can cause direct light detection issues\cite{Golwala:2012}.}\brep{For KIDs, which now primarily use single-layer interdigitated capacitors (IDCs), such a limitation leads to undesirable detector characteristics such as sensitivity to light unintentionally routed by the optically inactive capacitor to the inductor\cite{Golwala:2012} and/or a large capacitor footprint.}
\rsout{The use of amorphous dielectrics would solve these problems by allowing the fabrication of multi-layer components like PPCs.}\brep{Parallel-plate capacitors (PPCs) enabled by amorphous dielectrics would eliminate these properties.}
\mnote{The a-Si:H in Jonas' review is from Mazin:2010, so I removed Jonas ref and put that in.}
\rsout{However, previously published measurements show noise levels more than a 100 times higher in amorphous dielectrics compared to the lowest noise achieved in crystalline dielectrics\cite{Mazin:2010,Kouwenhoven:2024,Sun:2024}.}\brep{However, literature measurements of superconductive resonators using amorphous dielectrics\cite{Gao:2008,Zmuidzinas:2012,Kouwenhoven:2024,Sun:2024} show TLS noise levels more than 100 times larger than observed on crystalline substrates.}
In this article, we report on 
\rsout{the development of} 
\brep{superconductive resonators using PPCs that incorporate}
hydrogenated amorphous silicon (a-Si:H) with extremely low TLS noise, comparable to that of \brep{resonators fabricated on} crystalline silicon and sapphire.  
Because it is easily depositable, this novel a-Si:H 
\rsout{is crucial for the}
\brep{opens the door to} 
development of low-noise superconductive devices with multi-layer architectures.  

\rsout{In addition to noise} \brep{As expected from the fluctuation-dissipation theorem}, TLS\rsout{s} are \rsout{also shown to generate}\brep{observed to contribute} loss at low temperature, limiting the quality factor of superconductive resonators, the transmission of microstriplines, and \brep{the} coherence of qubits\cite{Martinis:2005, Gao:2008b, OConnell:2008, Mazin:2010, Bruno:2012, Buijtendorp:2020, Hahnle:2021, Buijtendorp:2022, Kouwenhoven:2024}.  We have shown in a previous article\cite{Defrance:2024} that our a-Si:H \rsout{has}\brep{recipes have} a radio-frequency (RF) loss tangent at 0~K and low \rsout{readout}\brep{stored} power of $7 \times 10^{-6}$, which is the lowest published loss tangent for amorphous dielectrics\rsout{,} and 
\rsout{comparable to}\brep{approaches} that of crystalline silicon\brep{\cite{Weber:2011}}.  
The TLS noise results \rsout{presented in this article}\brep{we present here} were obtained with the same a-Si:H recipes and the same devices as those \rsout{presented in our article on low TLS loss\cite{Defrance:2024}}\brep{loss results}.  All the fabrication details for the a-Si:H recipes can be found \rsout{in this previous publication}\brep{there}.

\rsout{For each test device,} \brep{Each device comprises} a 50 $\Omega$ \rsout{readout}coplanar waveguide (CPW) \brep{readout} feedline \rsout{is}inductively coupled to six lumped element LC resonators.  These LC resonators, made with Niobium (Nb), are composed of an inductor and two PPCs in series, with a 800~nm layer of a-Si:H \rsout{in between the capacitor plates} \brep{as the capacitor dielectric}.  The resonance frequencies of the six resonators are \rsout{gathered in two groups,}\brep{grouped into two triplets} \rsout{respectively} centered on 0.84~GHz and 1.55~GHz, and\brep{,} within each \rsout{group}\brep{triplet,} the designed resonance frequencies are \rsout{separated by a 5\% frequency difference}\brep{in the ratio 1:1.05:1.10}. 
Two different a-Si:H recipes (designated with the letters A and B) were used to fabricate these devices.  For each recipe, \rsout{a batch of 4 devices was fabricated and each device from a same batch is}\brep{one wafer with 4 devices was fabricated, with the devices from a given wafer} designated with a number from 1 to 4.  The fabrication was carried out
\rsout{by A. Beyer}at the Caltech Kavli Nanoscience Institute (KNI) clean room using PECVD for recipe A\rsout{,} and at the NASA Jet Propulsion Laboratory’s MicroDevices Laboratory (MDL) using inductively coupled plasma \brep{(ICP)} PECVD for recipe B.  

\mnote{Need to be clear about feedline readout power vs.\ stored power.  Use $P_{feed}$ and $P_{res}$; $P_{read}$ is ambiguous.  Use ``feedline'' in front of ``readout power'' also.}
Two TLS noise measurement campaigns were conducted.  At Caltech\brep{,} we measured the TLS noise as a function of \brep{feedline} readout power for five devices at a base temperature of 230~mK.  At NIST (National Institute of Standards and Technology), using a dilution fridge able to reach temperatures as low as 20~mK, we measured the TLS noise as a function of power and temperature for two of the five devices initially measured at Caltech.  

Caltech's experimental setup is the same as described \rsout{in Defrance et al.}\brep{previously}\cite{Defrance:2024}\rsout{,} except that we replaced the VNA (vector network analyzer) by the readout system developed by Minutolo et al.~\cite{minutolo:2019} using a USRP~X300 (universal software radio peripheral) instrument with UBX-160 daughterboard, commercialized by the company Ettus \footnote{https://www.ettus.com/}.  
\rsout{At the difference of the VNA, the USRP is able to read multiple frequency tones at a rate up to 200~MHz, which is useful for noise measurement.}\brep{Unlike a VNA, this system records the on-resonance network transmission time-stream, which we use to measure its noise power spectral density.  The system's 160~MHz RF bandwidth, centered on a local oscillator whose frequency can be set between \FD{DC} and \FD{6~GHz}, enables noise PSD measurements for multiple resonators.}
To measure TLS noise at different \brep{resonator stored} powers/energies \rsout{without saturating the USRP or be dominated by digitization noise}\brep{while maintaining constant USRP output and input power levels and thus signal-to-noise ratio}, \rsout{the readout tones go through} room\brep{-}temperature variable attenuators before and after the \rsout{device.} \brep{cryostat are used:}
\rsout{The}\brep{the} \brep{feedline} readout power at the device is swept by changing the value of the input attenuator, and the output attenuator is varied by \rsout{the opposite}\brep{a cancelling} amount\brep{.} \rsout{to keep the power read by the USRP constant.}

\mnote{I moved this from after the calibration and PSD calculation discussion because those are the same for both setups.}\rsout{Instead of relying on a USRP,}\brep{The} NIST \brep{setup instead} uses a \rsout{more common IQ}\brep{standard} homodyne \brep{IQ mixing}\rsout{test} setup. 
A signal generator emits a tone at $f_{res}$ that \rsout{goes through} \brep{is routed to} the device, \rsout{is}amplified by a \brep{cryogenic} low-noise \rsout{HEMT (}high electron mobility transistor\rsout{)} \brep{(HEMT)} amplifier, and
\rsout{is compared to}\brep{mixed with} the original signal by a TSC AD0540 IQ mixer, as described \rsout{in  Gao et al.}\brep{previously}\cite{Gao:2007}.  An \rsout{ADC (}analog-to-digital converter\rsout{)} \brep{(ADC)} digitizes the \rsout{I (}in-phase\rsout{)} and \rsout{Q (}quadrature\rsout{-phase)} \brep{(I and Q)} signals generated by the IQ mixer\rsout{,}. \rsout{and \rsout{like}}\brep{As} for the Caltech setup, variable attenuators before and after the \rsout{device}\brep{cryostat} ensure that the power \brep{levels} received by the IQ mixer and the ADC stay\rsout{s}  constant \brep{as the power at the device is varied}.

\begin{figure}[t]
\includegraphics[width=7cm]{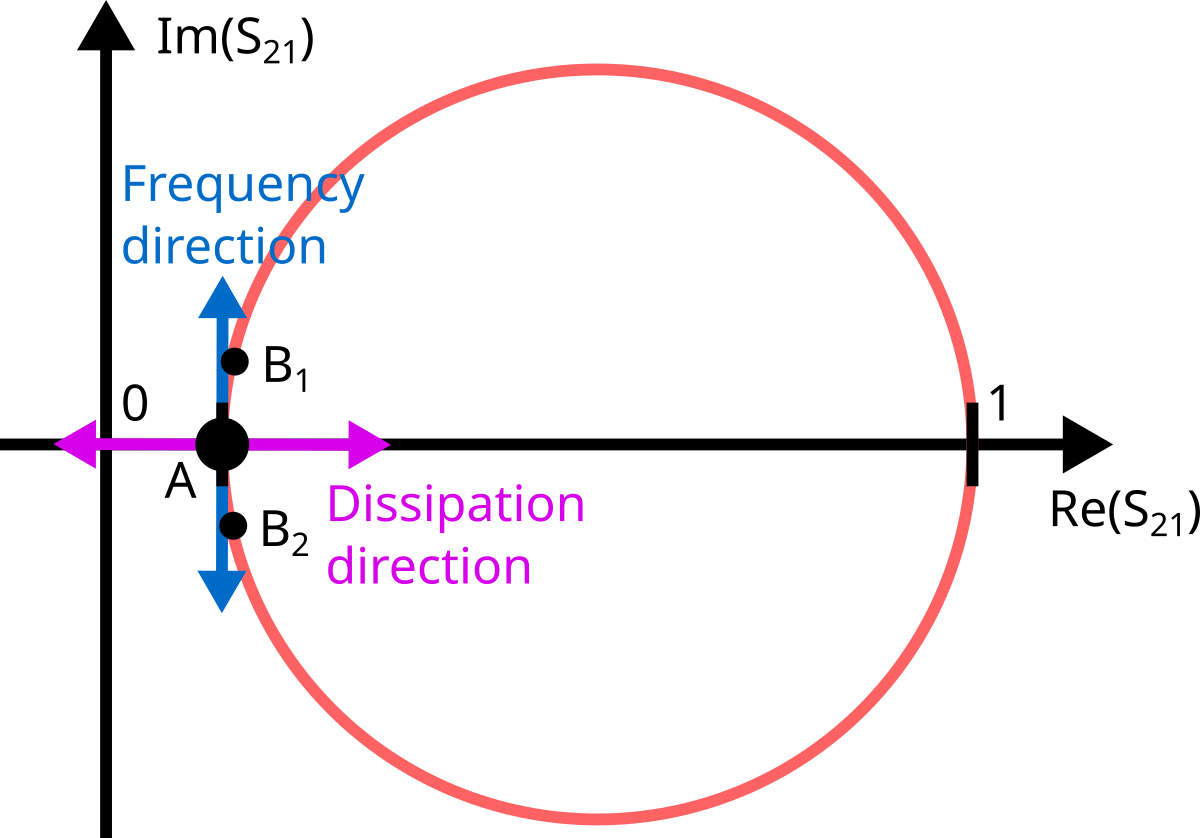}
\caption{\label{fig:IQ_circle} After removal of the cable delay component, the \protect\brep{complex} transmission \protect\rsout{($S_{21}$)} of the resonator as a function of frequency\protect\brep{, $S_{21}(f)$,} follows the red circle (also called \protect\brep{the ``}resonance circle\protect\brep{''} or \protect\brep{``}IQ circle\protect\brep{''})\protect\rsout{, when plotted in the complex plane}.  \protect\brep{The orientation displayed may require removal of rotations of the circle about the complex origin and/or about its own center\cite{Khalil:2012}.}  The point A on the circle, located at $f_{res}$, represents the location of the noise measurement.  The two points \protect\rsout{$B_1$}\protect\brep{B\textsubscript{1}} and \protect\rsout{$B_2$}\protect\brep{B\textsubscript{2}}, on each side of A, show the calibration measurements \protect\brep{that identify the frequency-dissipation basis}.}
\end{figure}

Each noise measurement includes three steps:
\begin{enumerate}
    \item Resonance fit: we measure the \brep{complex} transmission \rsout{$S_{21}$} as a function of frequency\brep{, $S_{21}(f)$,} across the resonance.  Using the python package SCRAPS\cite{Carter:2017}, we fit \rsout{the}$S_{21}$\brep{$(f)$} and extract the resonance frequency $f_{res}$, the internal quality factor $Q_i$\brep{,} and coupling quality factor $Q_c$.  
    \item Calibration: \brep{$S_{21}(f)$} is measured for two tones \rsout{are recorded}a few tens of kHz \rsout{before and after}\brep{below and above} $f_{res}$ (\rsout{$B_1$}\brep{B\textsubscript{1}} and \rsout{$B_2$}\brep{B\textsubscript{2}} in Figure\rsout{.}~\ref{fig:IQ_circle}).  The direction \rsout{$\overrightarrow{B_1 B_2}$}\brep{$\overrightarrow{\text{B}_{\text{1}} \text{B}_{\text{2}}}$} corresponds to the frequency direction 
    \rsout{of the noise} (tangent to the \rsout{IQ}\brep{resonance} circle), while the distance \rsout{$|B_1 B_2|$}\brep{$|\text{B}_{\text{1}} \text{B}_{\text{2}}|$}, in volts, \rsout{is used to calculate}\brep{provides} the volts-to-fractional-frequency \brep{conversion} coefficient $\zeta = \Delta f / (\rsout{|B_1 B_2|}\brep{|\text{B}_{\text{1}} \text{B}_{\text{2}}|} \times f_{res})$, with $\Delta f$ the frequency \rsout{distance}\brep{difference} between \rsout{$B_1$}\brep{B\textsubscript{1}} and \rsout{$B_2$}\brep{B\textsubscript{2}}.  
    \item Noise measurement: time-streams of 60~seconds are recorded\brep{.}
    \rsout{with the USRP at}
    \rsout{a rate of 200~MHz at} \rsout{$f_{res}$.}
\end{enumerate}

\begin{figure}[t]
\includegraphics[width=8.5cm]{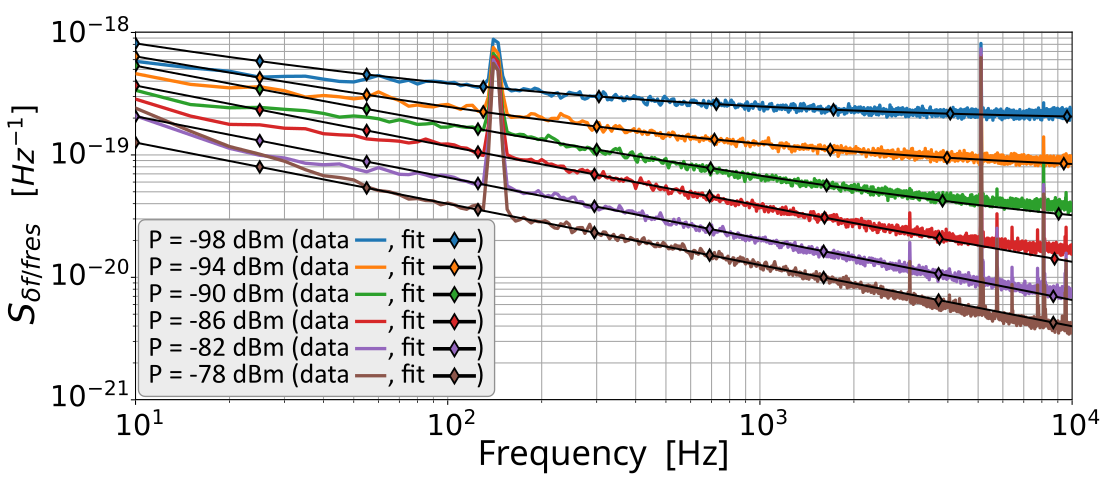}
\caption{\label{fig:fit_Sdff} Fractional\protect\brep{-}frequency noise PSD ($S_{\delta f/f_{\protect\brep{res}}}$) of \protect\brep{847~MHz resonance on} device B(2) \protect\rsout{resonance 847~MHz} as a function of audio frequency $\nu$.  The \protect\brep{feedline} readout power at the device was swept between $-98$~dBm and $-78$~dBm, corresponding to about 200 -- 2000~V/m in electric field\protect\rsout{,} and 3$\times 10^{6}$ -- 3$\times 10^{8}$ in photon number. $S_{\delta f/f_{\protect\brep{res}}}$ \protect\rsout{is mainly composed of white noise and TLS noise that can be fitted by $a \times f^{-0.5} + b $, with $a \times f^{-0.5}$ the TLS noise and $b$ the white noise.}
\protect\brep{can be fit well by a sum of TLS and white noise, $a \times f^{-0.5} + b$.}}
\end{figure}

\rsout{The recorded noise data is decimated down}\brep{The Caltech USRP setup provided noise time-streams sampled at 2~MHz,} \rsout{to filter out the high frequency noise (the decimation function also includes a low pass filter to avoid aliasing),}\brep{while the NIST setup provided time-streams sampled at 2.5~MHz, both after appropriate anti-alias filtering.} \rsout{and the}\brep{The} cable delay, obtained from the resonance fit, is removed.
\rsout{This}\brep{Using the frequency direction determination and the conversion coefficient $\zeta$ from the calibration dataset, the}
\rsout{data is} \brep{noise time-streams} are \brep{rotated to the frequency-dissipation basis and the frequency-direction time-stream} converted \rsout{in}to fractional\brep{-}frequency units\brep{.}
\rsout{using the volts-to-fractional-frequency coefficient $\zeta$ obtained during the calibration run, split into dissipation noise and frequency noise.}
\mnote{Can drop discussion of dissipation direction, not used further.}
\mnote{Since we don't have commensurate information for NIST, let's just drop all this PSD frame size business.  It is not important, and our readers know how to calculate PSDs.}\rsout{For each direction (frequency and dissipation), the 60~second long noise data is are split into 30 frames, 2~seconds long each, and the PSDs (power spectral densities) of all the frames are averaged using the Welch method to reduce the random noise and smooth the PSD data.}
\rsout{We finally obtain a} 
\brep{Fractional-}frequency noise PSD\brep{s} \rsout{in fractional frequency units that is primarily composed} \brep{are then calculated.  They} consist \brep{primarily} of white noise and TLS noise, as shown in Figure~\ref{fig:fit_Sdff}.  

To measure the TLS noise as a function of \rsout{readout}\brep{stored} power, \rsout{the} devices A(1), A(2), B(1), B(2), and B(3) were tested \rsout{at Caltech inside a cryostat} \brep{in the Caltech setup} at 230~mK.  Across these five devices, a total of 18 resonances were measured at \rsout{different}\brep{a range of feedline} readout powers, \rsout{between about}\brep{approximately} $-$100~dBm to $-$75~dBm (at the device). 
The \brep{fractional-}frequency noise PSD measurements and corresponding fits for \brep{the 847~MHz resonance on} device B(2)\rsout{, resonance 847~MHz,} are presented in Figure~\ref{fig:fit_Sdff}. These measurements, conducted at multiple \brep{feedline} readout powers, illustrate the characteristic TLS noise behavior observed \rsout{across}\brep{for} most resonances.
The main parasitic \rsout{peaks}\brep{spectral lines} seen in Figure~\ref{fig:fit_Sdff} come from the \rsout{pulse tube valve motor which is attached to the cryostat}\brep{cryostat's pulse-tube cooler valve motor}.  
\rsout{As we need the pulse tube on to ensure a stable temperature, and because the spikes did not prevent us from measuring the TLS noise, we left the pulse tube on while taking noise data.}\brep{Because these spectral lines did not impact the measurements or fitting, we did not shut off the pulse-tube cooler to eliminate them.}
\rsout{Measured}\brep{The measured} frequency noise PSDs are very reproducible above 100~Hz audio frequency, but\brep{, at lower frequencies,}\rsout{below 100~Hz,} they tend to vary slightly \rsout{from run to run and are}\brep{between datasets and cooldowns and are thus} less reliable. 
We have therefore \rsout{fitted the TLS noise within the frequency range where it was reliable and reproducible, from 200~Hz to 2~kHz (also avoiding the pulse tube 140~Hz peak).}\brep{restricted the fitting of a noise model to the range 200~Hz to 2~kHz.} 
\rsout{The frequency noise PSD data was fitted with the equation}\brep{The model consists of the sum of a TLS and a white noise term,} $a \times \nu^{\alpha} + b$, where \rsout{S\textsubscript{TLS} = $a \times \nu^{\alpha}$ is the TLS noise PSD and $b$ the white noise} \brep{$\nu$ is the audio frequency}.

Previous studies \cite{Burnett:2013,Burnett:2014,Gao:2007,Kouwenhoven:2024,Barends:2009,Gao:2008c} show that \brep{the TLS noise PSD}, \rsout{S\textsubscript{TLS}}\brep{$S_{\text{TLS}}$,} scales as \FD{$T^{-\beta}$}, $\rsout{\sim} P_{\rsout{read}\brep{res}}^{-0.5}$\brep{,} and $\rsout{\sim} \nu^{\alpha}$ \rsout{($\alpha \in [-1\rsout{;}\brep{,} -0.5]$),} with \brep{$T$ the resonator temperature,} $P_{\rsout{read}\brep{res}}$ the \rsout{circulating}\brep{stored} power in the resonator\brep{,} and $\nu$ the audio frequency.  
\mnote{Move our results to later in paper.  This section is literature review.}
\FD{For $T > 100$~mK, \rsout{the measured} \brep{literature} values of $\beta$ are: 2~\cite{Gao:2008}, 1.73~\cite{kumar:2008}, \rsout{[1.3, - 1.65]} \brep{1.3 -- 1.65}~\cite{Ramanayaka:2015}, \rsout{[1.2. - 1.4]} \brep{1.2 -- 1.4}~\cite{Burnett:2014}, $\approx$~0.79~\cite{Sun:2024}, and \rsout{[0.55, - 1.35]} \brep{0.55 -- 1.35}~\cite{Kouwenhoven:2024} (estimated graphically from Figure 11).  
The large range of $\beta$ can be explained by its power dependence.  Gao et al.\cite{Gao:2008} (Figure~5.21) and Kouwenhoven et al.\cite{Kouwenhoven:2024} 
show that $\beta$ decreases \rsout{when the readout power increases}\brep{with increasing stored power}, suggesting that \rsout{a} partial saturation of TLSs by \rsout{the readout}\brep{stored} power reduces the temperature dependence of TLS noise.} 
\rsout{Our measurements confirm this behavior with values of $\beta$ varying from 0.4 at the highest readout power, to 1.3 at the lowest power.}\brep{With regard to the $\nu$ dependence,} \rsout{While}\brep{while} most TLS noise measurements using homodyne setups find values of $\alpha$ close to $-0.5$\cite{Gao:2007,Barends:2009,Gao:2008c,Sun:2024}, measurements from Burnett et al.\cite{Burnett:2013,Burnett:2014} using a Pound lock\brep{ing} loop seem to indicate that\brep{,} at frequencies \rsout{lower than}\brep{below} about 10~Hz\brep{,} the \brep{logarithmic} slope of \rsout{S\textsubscript{TLS}}\brep{$S_{\text{TLS}}$} \rsout{follows}\brep{is} $\alpha = -1$, as expected for flicker frequency noise, but they did not measure $\alpha$ above 100~Hz.
\rsout{Lately,}\brep{Recently,} Kouwenhoven et al.\cite{Kouwenhoven:2024} were able to measure \rsout{S\textsubscript{TLS}}\brep{$S_{\text{TLS}}$} between 1~Hz and 10~kHz.  They found $\alpha \approx -1$ below 100~Hz and $\alpha \approx -0.5$ above 100~Hz -- 1~kHz, with a smooth transition between \rsout{both}slopes, which reconciles all previous $\alpha$ results.  They also show that the \rsout{two $\nu^{-1/2}$}\brep{$\nu^{-0.5}$} and $\nu^{-1}$ \rsout{parts of the curve}\brep{model components} follow the \rsout{same typical}\brep{usual} TLS power and temperature dependence\brep{s}.  
We note that the presence of both slopes had also been seen by Gao et al. in 2007\cite{Gao:2007}, with a transition around 10~Hz, but\brep{,} at the time\brep{,} the $\alpha = -1$ slope had been attributed to readout electronics noise.
Our measurements and fits of \rsout{S\textsubscript{TLS}}\brep{$S_{\text{TLS}}$} between 200~Hz and 2~kHz follow $\alpha \approx -0.5$, in accordance with previous measurements at $\nu > $100~Hz.
Our most reliable fits of $\alpha$ were obtained at high \brep{feedline} readout powers\brep{,} where the white noise level is low compared to the TLS noise\rsout{,} and the TLS noise slope \brep{is} clearly visible, as illustrated by Figure~\ref{fig:fit_Sdff}.
To reduce the degeneracy between the \rsout{three noise variables,}\brep{fit parameters} $a$, $b$, and $\alpha$, and because $\alpha$ is \rsout{assumed to stays constant across}\brep{expected to be the same for all} resonators, we used the value \rsout{of}$\alpha = -0.5$ obtained from the fits at high \brep{feedline} readout powers as a fixed parameter for all the \rsout{resonances}\brep{fits}, only \rsout{fitting}\brep{varying} $a$ and $b$. 

\begin{figure}[t]
\includegraphics[width=8.5cm]{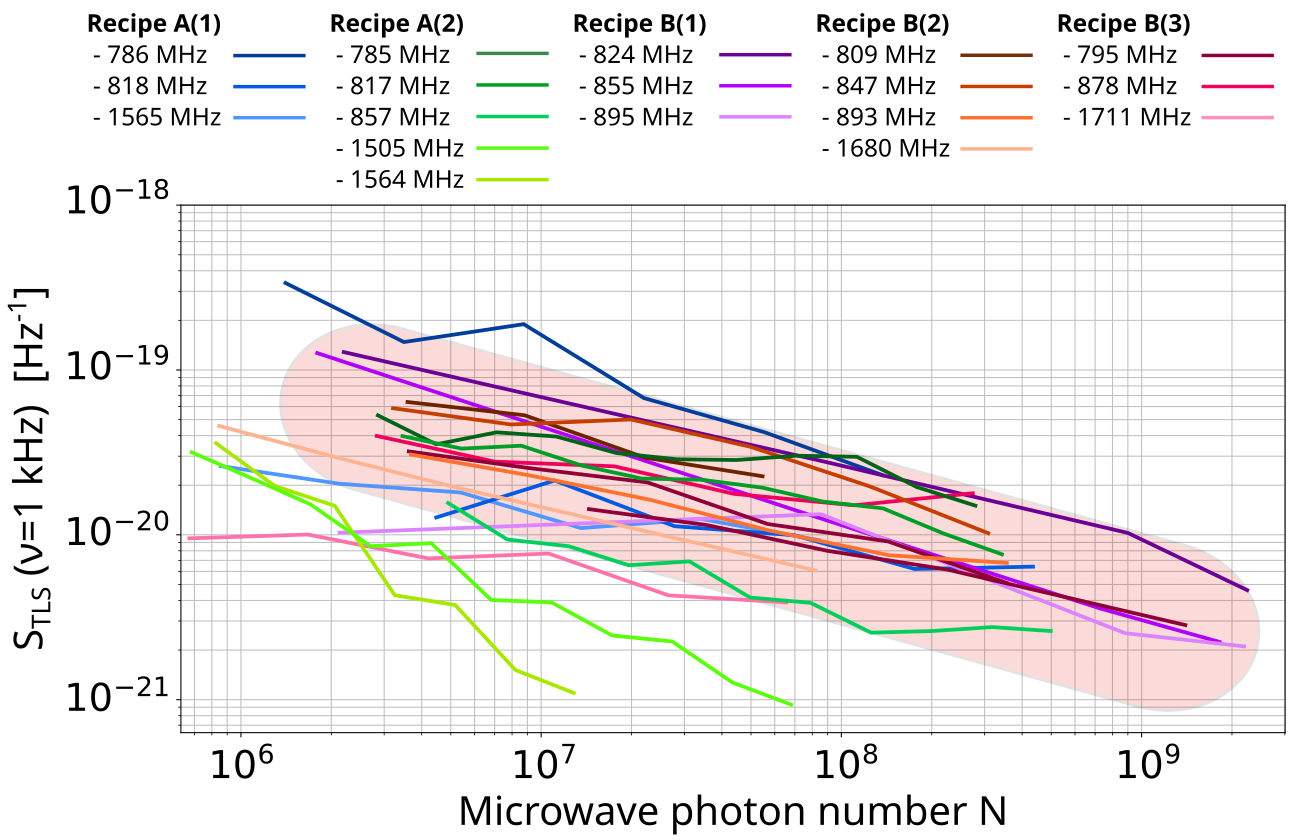}
\caption{\label{fig:TLS_1kHz_Caltech} TLS fractional\protect\brep{-}frequency noise \protect\brep{PSD} measured at 1~kHz \protect\rsout{for a variety of resonators, plotted} as a function of \protect\rsout{the} stored microwave energy measured in photon units\protect\rsout{,} for \protect\brep{18 resonators across} five devices fabricated using recipes A and B.}
\end{figure}

Using the values of $a$ given by the fit, we \rsout{calculate}\brep{show in Figure~\ref{fig:TLS_1kHz_Caltech}} \rsout{S\textsubscript{TLS}}\brep{$S_{\text{TLS}}$} at $\brep{\nu = }1$~kHz\rsout{, S\textsubscript{TLS} ($\nu = $1~kHz) = $a \times \nu^{-0.5}$,} as a function of stored microwave energy, expressed in photon number, for all the devices measured at Caltech. \rsout{The results are shown in Figure~\ref{fig:TLS_1kHz_Caltech}.}For each resonator, \rsout{using their coupling quality factor $Q_c$, total quality factor $Q_r = 1/(Qc^{-1} + Qi^{-1})$, resonance frequency $f_{res}$, and the Plank constant $h$, we calculated the microwave photon number corresponding to each readout power $P_{read}$}\brep{we calculated the microwave photon number corresponding to each feedline readout power $P_{feed}$ using}:
\begin{equation}
    \label{eq:photon_N}
    N = \frac{W}{h f_{res}} = \brep{\frac{P_{res}}{h f_{res}^2}} = \frac{P_{\rsout{read}\brep{feed}}}{\pi h f_{res}^2} \frac{Q_r^2}{Q_c}.
\end{equation}
\brep{where $P_{res}$ is the stored power in the resonator and $h$ is Planck's constant.}
\rsout{More information about how equation~\ref{eq:photon_N} was obtained and the correspondence to electric field is given in Defrance et al. \cite{Defrance:2024}.}\brep{The provenance of Equation~\ref{eq:photon_N} and the correspondence to electric field are available elsewhere\cite{Defrance:2024}.}
\brep{We see some variation \rsout{A}a}mong the 18 curves shown in Figure~\ref{fig:TLS_1kHz_Caltech}, \rsout{we see some disparity between them, }even for \rsout{measurements coming from}\brep{resonators incorporating the same a-Si:H film on} the same device.  
Multiple measurements were taken for each resonator and they are reproducible, suggesting that the variations \rsout{might come}\brep{may arise} from \rsout{fabrication differences}\brep{film non-uniformity} or defects \rsout{: surface cleaning in-homogeneity or adhesion problems between layers could be responsible for TLS density variation across the device, leading to slightly different TLS noise levels for each resonator.}\brep{leading to TLS density variations.}
Figure~\ref{fig:TLS_1kHz_Caltech} \rsout{does not} show\brep{s no}\rsout{any} 
obvious \rsout{S\textsubscript{TLS}}\brep{difference in $S_{\text{TLS}}$} between recipes A and B\rsout{, suggesting that both a-Si:H recipes give similarly low TLS noise, which might be consistent with the fact that both recipes were also giving similarly low TLS loss tangents\cite{Defrance:2024}}. \brep{The 1 decade variation in $S_{\text{TLS}}$ in Figure~\ref{fig:TLS_1kHz_Caltech} combined with the modest factor of $\approx$2 difference in loss tangent between the recipes would, however, hide all but very strong dependences of noise level on loss tangent.} \rsout{For a better visualization}\brep{To give a sense of the range} of a-Si:H TLS noise level \brep{for our recipes}, we \rsout{have enclosed}\brep{define in Figure~\ref{fig:TLS_1kHz_Caltech} a red envelope inside of which} most of the \rsout{S\textsubscript{TLS} vs. N}\brep{$S_{\text{TLS}}$ vs.\ $N$} \rsout{measurements in a red area}\brep{reside}, \rsout{only}leaving out \brep{only} three outlier \rsout{curves having a noise level and slope very different from other curves}\brep{datasets}.

\begin{figure}[t]
\includegraphics[width=8.5cm]{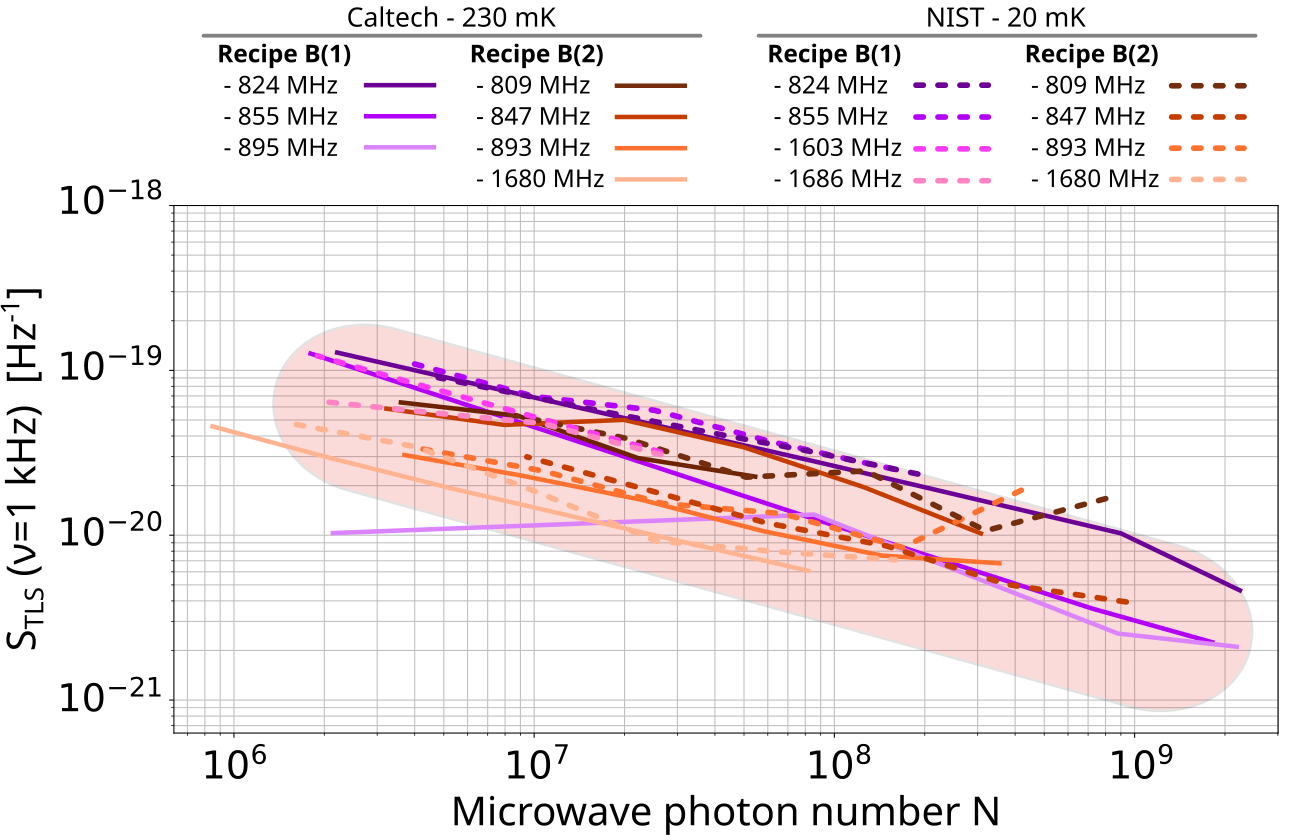}
\caption{\label{fig:TLS_1kHz_NIST} TLS fractional\protect\brep{-}frequency noise PSD measured at 1~kHz for devices B(1) and B(2) at Caltech (230~mK) and NIST (20~mK).}
\end{figure}

To \rsout{confirm}\brep{check} the results obtained at Caltech\rsout{ with the USRP system}, we re-tested devices B(1) and B(2) at NIST three years later. 
The TLS fractional\brep{-}frequency noise PSDs at $\nu = 1$~kHz measured at NIST at 20~mK are plotted in Figure~\ref{fig:TLS_1kHz_NIST}\rsout{ and}\brep{,} overlaid with Caltech \rsout{USRP} measurements obtained for the same devices at 230~mK.  For device B(1), not all resonances gave reliable TLS noise \rsout{results}\brep{measurements} with both systems.  \brep{The} Caltech data for \brep{the} \rsout{resonances} 1603~MHz and 1680~MHz \brep{resonances} \rsout{was showing a}\brep{displayed} white noise well above \brep{the expected} TLS noise level\rsout{, preventing any reliable TLS noise fit, and the}\brep{. The} NIST data for \brep{the}\rsout{resonance} 895~MHz \brep{resonance} \rsout{was showing}\brep{evidenced} a strong additional noise \rsout{that we did not manage to clean, preventing us from extracting and fitting the TLS noise}\brep{that rose steeply with decreasing audio frequency}.
Comparing the six \brep{remaining} resonances \rsout{that are}common \rsout{between}\brep{to the} Caltech and NIST \rsout{noise}data\brep{sets}, we \rsout{see that there is}\brep{observe} less than a factor 2 difference\rsout{ between measured S\textsubscript{TLS} data}.

\rsout{While it is reassuring to obtain such similar results from measurements taken 3 years apart with completely different test setups, we would ideally have expected S\textsubscript{TLS} to be about 1.5 to 2 times higher in NIST measurements (taken at 20~mK) than in Caltech data (taken at 230~mK), as suggested by the temperature dependence of TLS noise, T\textsuperscript{-0.8} -- T\textsuperscript{-1.3}, observed above 100~mK for energies below $N = 10^8$ in these devices.}\brep{While the consistency between datasets is encouraging, we would have expected the NIST measurements to be higher than the Caltech data due to the temperature dependence of TLS noise.} \mnote{Moved our $T$ dependence from literature review.}\brep{In data taken at NIST not presented here, we found consistency with the literature temperature dependence reviewed earlier, with values of $\beta$ varying from 0.4 to 1.3 over the stored powers probed.  For $N \lesssim 10^8$, these dependences imply an expected factor of 1.5~--~2 difference between the NIST and Caltech datasets.} 
\rsout{Since the readout power at the device is known with an accuracy of about 2-3~dBm for each test setup, the most likely explanation for the similar S\textsubscript{TLS} levels is an effective readout power higher by a few dBm in the NIST setup, and/or lower by a similar amount in the Caltech setup.}\brep{The most likely explanation for this apparent discrepancy is a systematic error in the feedline readout power and thus the microwave photon number: the feedline readout power at the device is only known to an accuracy of 2~--~3~dBm.}

\begin{figure}[t]
\includegraphics[width=8.5cm]{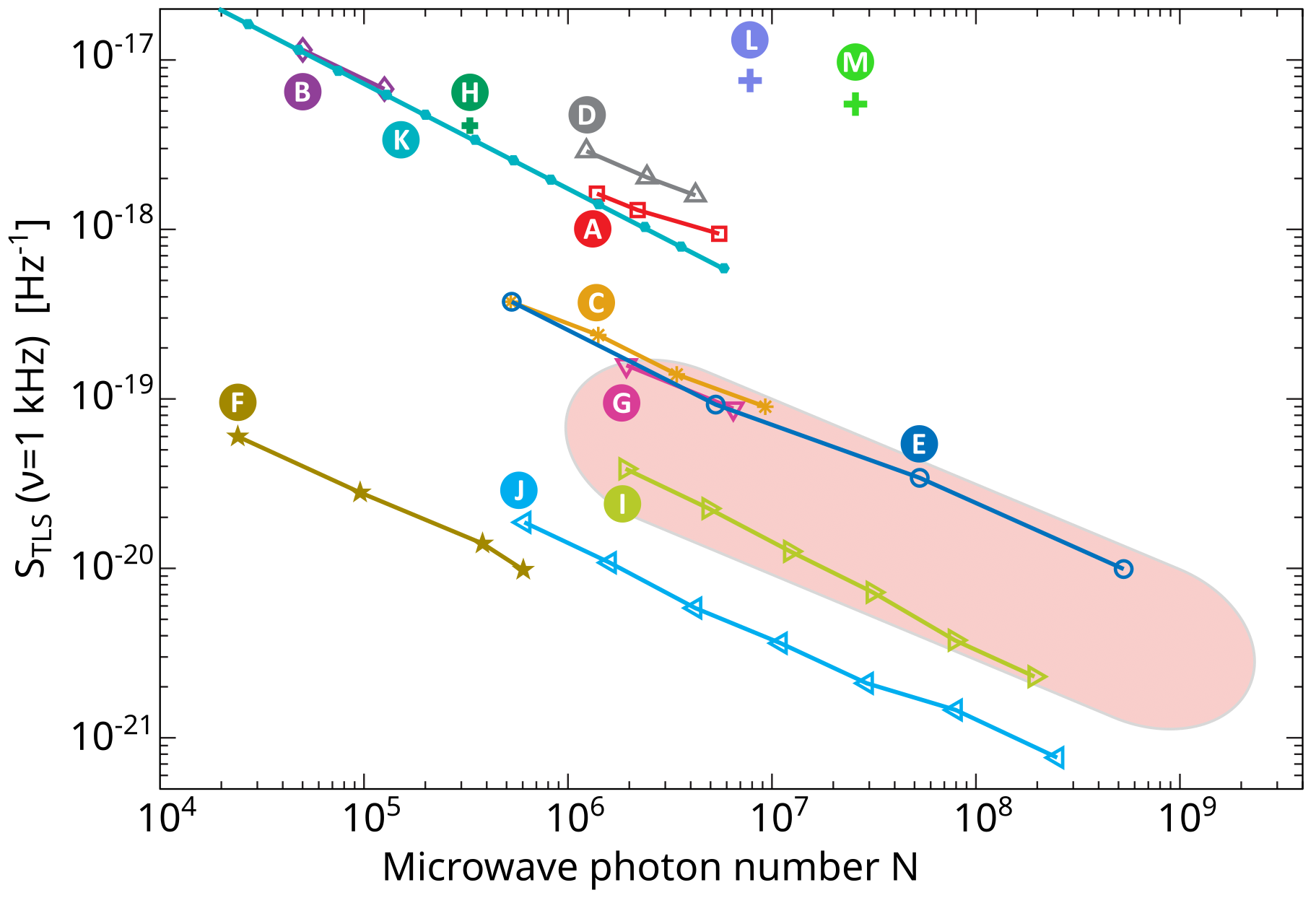}
\caption{\label{fig:TLSnoise_jonas} TLS fractional\protect\brep{-}frequency noise \protect\brep{PSD} measured at 1~kHz \protect\rsout{for a variety of resonators plotted} as a function of \protect\rsout{the} stored microwave energy measured in photon units \protect\brep{for a variety of resonators}.  (A) Al on Si, $f_{res}$= 5.8~GHz, T=120~mK\cite{Gao:2007}; (B) Al on Si \protect\brep{CPW}, 4.8~GHz, 120~mK\cite{Gao:2007}; (C) Al on sapphire \protect\brep{CPW}, 4~GHz, 120~mK\cite{Gao:2007}; (D) Al on Ge \protect\brep{CPW}, 8~GHz, 120~mK\cite{Gao:2007}; (E) Nb on Si \protect\brep{CPW}, 5.1~GHz, 120~mK\cite{Gao:2007}; (F) Nb IDC on Si with Al CPW inductor, 5.6~GHz, 120~mK\cite{Noroozian:2009}; (G) TiN on Si \protect\brep{CPW}, 6~GHz, 100~mK\cite{Leduc:2010}; (H) \protect\brep{Al-a-Si:H-Al} microstrip \protect\rsout{Al on a-Si:H + Al ground plane}, 9~GHz, 150~mK; (I) NbTiN on Si \protect\brep{CPW}, 4.4~GHz, 310~mK\cite{Barends:2010}; (J) NbTiN on Si \protect\brep{CPW}, 2.64~GHz, 310~mK\cite{Barends:2010}; (K) NbTiN\protect\rsout{ on }\protect\brep{-}a-Si:C\protect\brep{-NbTiN PPC}, 5.15~GHz, 100~mK\cite{Kouwenhoven:2024}; (L) Al\protect\rsout{ on }\protect\brep{-}SiN\textsubscript{x}\protect\rsout{ with }\protect\brep{-}Al \protect\rsout{ground plane}\protect\brep{PPC}, 1.9~GHz, 100~mK\cite{Sun:2024}; (M) Al\protect\rsout{ on }\protect\brep{-}Si\textsubscript{3}N\textsubscript{4}\protect\rsout{ with }\protect\brep{-}Al \protect\rsout{ground plane}\protect\brep{PPC}, 2.2~GHz, 100~mK\cite{Sun:2024}. 
\textbf{(Red \protect\rsout{area}\protect\brep{envelope}) Nb\protect\rsout{ on }\protect\brep{-}a-Si:H\protect\rsout{ with }\protect\brep{-}Nb \protect\rsout{ground plane}\protect\brep{PPC}, 800~MHz and 1.6~GHz, 20~mK and 230~mK.} 
Adapted with permission from J.~Zmuidzinas~\cite{Zmuidzinas:2012}.  Copyright 2012, Annual Reviews.
}
\end{figure}

In order to compare our a-Si:H TLS noise results with the literature, we overlaid \rsout{the red area corresponding to our a-Si:H results}\brep{the envelope of our measurements on Figure~14 of} \rsout{on J.~Zmuidzinas}\brep{the} review paper \brep{by Zmuidzinas}\cite{Zmuidzinas:2012}\rsout{ plot}, reproduced in Figure~\ref{fig:TLSnoise_jonas} \brep{and augmented with recent results on other amorphous dielectrics (K, L, and M)}.
\rsout{To extrapolate the K data from 10~Hz to 1~kHz, we used $\alpha$ = -1, consistent with the scaling observed by Kouwenhoven et al. in their original dataset.}
\brep{The K data were reported at $\nu = 10$~Hz, so we used their measurement of $\alpha \approx -1$ to extrapolate to $\nu = 1$~kHz\cite{Kouwenhoven:2024}.}
We see that the \rsout{S\textsubscript{TLS}}\brep{$S_{\text{TLS}}$}($\nu = 1$~kHz) results presented in this article are 8 -- 80 times lower than previous measurements for a-Si:H (H), 5 -- 50 times lower than \brep{for} hydrogenated amorphous silicon carbide \brep{(}a-SiC:H\brep{;} (K)\brep{)}, >100 times lower than \brep{for} silicon nitride \brep{(}SiN\textsubscript{x} (L) \& Si\textsubscript{3}N\textsubscript{4} (M)\brep{)}, and comparable to the TLS noise level usually achieved \rsout{with}\brep{on} crystalline substrates\rsout{, like}\brep{ such as}\rsout{crystalline} silicon (A, B, E, F, G, I, J) and sapphire (C).

\mnote{I don't think we need this paragraph in such a short article.}\rsout{In conclusion, we have developed two recipes of a-Si:H and used them to fabricate five test devices (2 using recipe A and 3 using recipe B).  A total of 18 resonators, distributed over these devices, were measured at Caltech at 230~mK.  The fits of the TLS noise PSD between 200~Hz and 2~kHz allowed us to extract the TLS noise level at 1~kHz, plot it as a function of the energy stored in the resonators, and compare it with other published results.  The TLS noise level obtained for our 18 resonators is comparable to that of crystalline dielectrics, and >5 times lower than the lowest measured TLS noise in amorphous dielectrics to date.  These results were confirmed by measuring two of the 5 devices at NIST at 20~mK, with a different test setup.  While we still observe some TLS noise level variations among resonators, we believe that they are caused by fabrication inhomogeneities that we intend to correct in the future.  Even accounting for these variations, the exceptionally low level of TLS noise reached by our a-Si:H recipes is a key milestone to enable the emergence of a new generation of multi-layer low-noise superconductive resonators.}\brep{We conclude by noting that the competitive level of TLS noise provided by the two a-Si:H recipes reported here renders PPC-based architectures for superconductive resonators a viable alternative to those using CPWs or IDCs.  As noted earlier, such an architecture could be transformative for KIDs, vastly reducing their footprint and capacitor-routed direct optical absorption.  We are actively applying these a-Si:H PPCs in KIDs for mm/submm astronomy for continuum imaging (the NEW-MUSIC\cite{Golwala:2024} instrument for the Leighton Chajnantor Telescope) and filterbank spectroscopy\cite{Kovacs:2012}.}

\begin{acknowledgments}
This work has been supported by the JPL Research and Technology Development Fund, the National Aeronautics and Space Administration under awards 80NSSC18K0385 and 80NSSC22K1556, and the Department of Energy Office of High-Energy Physics Advanced Detector Research program under award DOE-SC0018126.  A. B. and F. D. carried out research/fabrication at the Jet Propulsion Laboratory, operated by the California Institute of Technology under a contract with the National Aeronautics and Space Administration (80NM0018D0004).  The authors would like to thank J.~Gao for his very helpful comments, \brep{D.~Cunnane for early discussions of a-Si:H fabrication techniques, }\brep{P.~Day for early measurements of a-Si:H PPC TLS noise,} and M.~Hollister for design and construction of the cryostat used for this work.
\end{acknowledgments}

\bibliography{biblio}

\end{document}